\documentclass[prl,aps,preprint]{revtex4}
\begin{document}
\setcounter{page}{1}
\date{\today}
\title{Anisotropy and percolation threshold in a multifractal support}

\author{   L. S. Lucena, J. E. Freitas, G. Corso, and R. F. Soares. } 

\affiliation{ International Center for Complex Systems and    
 Departamento de F{\'\i}sica Te\'orica e Experimental,      
 Universidade Federal do Rio Grande do Norte, Campus Universit\'ario   
 59078 970, Natal, RN, Brazil.}

\begin{abstract}
 
Recently a multifractal object, $Q_{mf}$, was proposed 
to study percolation properties in a multifractal support. 
The area  and the number of 
neighbors of the  blocks of $Q_{mf}$ show a non-trivial behavior.
The value of the probability of occupation at the percolation 
threshold, $p_{c}$, is a function of $\rho$, a parameter 
of $Q_{mf}$ which is related to its anisotropy. 
We investigate the relation between $p_{c}$ 
and the average number of neighbors of the blocks as well as the  
anisotropy of $Q_{mf}$. 

\end{abstract}

\maketitle

\section{1- Introduction}

Due to the work of many physicists, and particularly to the 
contributions of Dietrich Stauffer, percolation theory has 
became a powerful tool in Science describing phenomena in 
many areas as geology, biology, magnetism, or social phenomena 
 \cite{stauffer,perc3}. Despite the enormous success of 
percolation it has been a theory studied in a   
support (lattice) that has a single dimension. 
The only references relating percolation and multifractality 
concern to the multifractal properties of some quantities of the 
spanning cluster at the percolation threshold \cite{multperc1,multperc2}.

Recently, a model to study percolation in a multifractal was 
proposed \cite{last} in the literature. In fact, the authors have 
created an original multifractal object, $Q_{mf}$, and an efficient 
way to estimate its percolation properties. In this work we study in 
detail the method to estimate  $p_c$ for this multifractal and 
discuss the relation between  $p_c$, some topologic characteristics, 
and the anisotropy of $Q_{mf}$. 

The multifractal object we develop, $Q_{mf}$, is an intuitive generalization of 
the square lattice \cite{last}. Suppose that in the construction of the square 
lattice we use the following algorithm: take a square  of size $1$ and 
cut it symmetrically with  vertical and  horizontal lines. Repeat this process 
$n$-times; at the $n^{th}$ step we have a regular square
lattice with $2^n \times 2^n$ 
cells. The setup algorithm of $Q_{mf}$ is quite similar, the main difference 
is that we do not cut 
the square in a symmetric way. In section $2$ we 
explain in detail this algorithm. 

The development of $Q_{mf}$ has a twofold motivation. 
Firstly, there are systems like oil reservoirs  
that show multifractal properties \cite{Hermann}
 and are good candidates to be modeled by such 
object. Secondly, there is indeed a much more general scope: 
we want to study percolation phenomena 
in lattices that are not regular, but that 
are multifractal in the geometrical 
sense. It is important to know how site percolation transition  happens in 
 lattices in which the cells vary in size and also in the number of neighbors.

In this work we analyse some geometric and topologic properties  of  
the percolation cluster generated  on $Q_{mf}$ at the percolation threshold. 
 The paper is organized as follows: in 
 \S2 we present the process of construction 
of $Q_{mf}$ and the algorithm for the estimation of $p_c$,
in \S3 we show the numerical simulations concerning  the 
percolation threshold $p_c$ and the topologic properties of $Q_{mf}$; and  
 finally in \S4 we present our final remarks and comments.

\section{2- The Model}

In this section we show the process of building of the multifractal $Q_{mf}$ 
and the key concepts to estimate $p_c$. 
 We start with a square of linear size $1$ and a partition parameter
$0<\rho<1$. For reasons that will be clear
 later, $\rho = \frac{s}{r}$, where $s$ and 
$r$ are integers.  We call the multifractal built from this 
parameter $\rho$ as the "$\rho-Q_{mf}$", or "$(r,s)-Q_{mf}$".

The first step, $n=1$, consists of two sections of 
the square: a vertical and an horizontal. 
Initially the square is cut in two pieces of 
area $\frac{r}{s+r}= \frac{1}{1+\rho}$ and 
$\frac{s}{s+r}= \frac{\rho}{1+\rho}$ by a vertical line.
This process is shown in figure \ref{fig1}(a), where we use as an example 
$\rho = \frac{s}{r} = \frac{2}{3}$.
 The horizontal cut  in which we use the same partition $\rho$ 
 is shown in figure \ref{fig1}(b). 
The first partition of the square generates
four rectangular blocks: the largest one of area $(\frac{\rho}{1+\rho})^2$,
two  of area $\frac{\rho}{(1+\rho)^2}$ and the
smallest one of area $(\frac{1}{1+\rho})^2$. 
The difference between the largest area and the smallest one increases 
as $\rho \rightarrow 0$, moreover, the blocks get 
more and more stretched in this limit. 
Therefore $\rho$ measures the anisotropy of 
 $Q_{mf}$.

In the second step, $n=2$, we 
repeat the same process of vertical and horizontal 
sections as in step $1$. Generically we get $2^{2 n}$ blocks after 
the $n^{th}$-step. 
The partition process produces a set of blocks with a variety of areas. 
 We call a set of all elements with the same area as 
a $k$-set. 
 At the $n^{th}$-step of the algorithm the 
partition of the square in  blocks
 follows the binomial rule:
\begin{equation}
         A= \sum_{k=0}^n \> C_k^n \> \left(\frac{\rho}{1+\rho}\right)^k 
\left(\frac{1}{1+\rho}\right)^{n-k} = 
\left(\frac{\rho + 1}{\rho + 1}\right)^n = 1.
\label{bino2}
\end{equation}
The number of elements of a $k$-set is $C_k^n$.  
In reference \cite{last} we see that as $n \rightarrow \infty$ all the $k$-sets 
determine a monofractal whose dimension is $d_k = lim_{n \rightarrow \infty} \frac{log \> C_n^k \>
\> s^k \>r^{(n-k)} }{log \> (s+r)^\frac{n}{2} }$. In this limit, the ensemble of all
$k$-sets engenders the multifractal object $Q_{mf}$ itself. 

Figure \ref{fig2} shows a picture of $Q_{mf}$ 
for $\rho = \frac{1}{3}$ and $n=4$. We use 
the following code color: blocks of equal area have the same tonality. In 
other words, all the blocks of a same $k$-set 
share a common gray-tone. 
The general view of the object shows an anisotropic, heterogeneous lattice with 
a non-trivial topology. The anisotropy of $Q_{mf}$ will change the percolation 
threshold as we investigate in the next section. Before 
that we study in more detail the problem of percolation.

The main subject of this work is the study of percolation properties of  
$Q_{mf}$. To perform such a task we develop a percolation algorithm.  
The percolation algorithm for  $Q_{mf}$
start mapping this object into the
square lattice. The square lattice should be large enough that
each line segment of $ Q_{mf}$
coincides with a line of the lattice, this condition 
imposes that $\rho$ is a rational number 
(it means, $r$ and $s$ are integers). 
Therefore we consider that the square lattice is more
finely divided than $Q_{mf}$. In this way all blocks
 of the multifractal are
composed by a finite number of cells of the square lattice. 
To explain the percolation algorithm we suppose that  $Q_{mf}$
construction is at step $n$.
We proceed the percolation algorithm by choosing at
random one among the $2^{2 n}$ blocks of
$ Q_{mf}$ independent of its size or number of neighbors. 
Once a block is chosen all the  cells
in the square lattice corresponding to this block are considered as
occupied. Each time a block of $ Q_{mf}$ is chosen
the algorithm check if the occupied cells at the underlying square 
lattice are connected in such a way to form a spanning  percolation
cluster. The algorithm to check percolation is similar to the one used in
 \cite{ziff,freitas0,freitas2,freitas3}. 

\section{3- Numerical Results}
 
In this section we show numerical results concerning the percolation 
threshold. Before the numerics we introduce some definitions.  
We call lattice the square lattice underlying  $Q_{mf}$. 
Following the literature \cite{stauffer} we call $p$ the probability 
of occupation of 
a lattice site. $R_L$  is the probability that for a site occupation $p$ there exists a 
contiguous cluster of occupied sites which crosses completely  the square lattice 
of size $L$.  
$p_c$ is the probability of occupation at the percolation threshold. 
There are several ways \cite{ziff} to define $R_L$. We use two of them: 
$R_L^e$  is the probability that there exits a cluster crossing  
either the horizontal or the vertical direction, and $R_L^b$ is the 
probability that there exits a cluster crossing  
both directions. At the limit 
of infinite lattice size $R_L^e$ and $R_L^b$ converge to a common value 
for the square lattice case. Besides we call $p_c^e$ the value of 
$p_c$ estimated from $R_L^e$, it means, the average $p_c$ over lattices that 
percolate in one direction, the horizontal or the vertical. And, similarly, 
$p_c^b$ the value of $p_c$ estimated from $R_L^b$.

Figure \ref{fig3} illustrates the behavior of $p_c^e$ and $p_c^b$ for $7$ different 
lattice sizes. Two typical values of $\rho$ are used: $\rho = \frac{2}{3}$ 
(solid line) and  $\rho = \frac{1}{4}$ (dashed line). In \ref{fig3} 
(a) we show both $p_c$ versus 
$\frac{1}{L}$, the upper values corresponds to $p_c^b$ 
 and the bottom values to $p_c^e$. The lattice size, $L=(r+s)^n$, 
corresponds to  $4 \leq n \leq 10$, in both cases $r+s=5$.  The data of the 
upper branch of the figure  collapse into 
a single curve, diversly from the lower branch. In other words, 
the curves of $(2,3)-Q_{mf}$ and $(4,1)-Q_{mf}$ share roughly the same 
$p_c^e$ but diverse $p_c^b$. The reason for this feature is the strong 
anisotropy of  $(4,1)-Q_{mf}$ compared to the case 
$(2,3)-Q_{mf}$. The anisotropy of the percolation 
cluster does not affect $p_c$ in both  
 directions, but it affects $p_c$ in one direction.  
In fact, the percolation in both 
directions comes from an average over both directions, in such situation, 
any eventual anisotropy of the percolation cluster vanishes  
because of the average. 

An useful way to define  $p_c$ is to take the average 
value $p_{c_{ave}} = \frac{p_c^e + p_c^b}{2}$.  
In figure \ref{fig3}(b) we plot $p_{c_{ave}}$ versus $\frac{1}{L}$ 
corresponding to the same data figure \ref{fig3}(a). We observe in this figure that 
both curves converge to a saturation value that is not the same. 
The difference between the two cases is  related to the curve 
of $p_c^b$. The anisotropy (stretching of the blocks) due to  
$\rho$ implies an anisotropy in the percolation cluster. Such anisotropy 
determines that the percolation cluster does not have a 
correlation length independent of the direction. Because the percolation 
cluster is anisotropic the symmetry between $p_c^e$ and $p_c^b$ should 
fail. Therefore, we expect that for smaller $\rho$ (more anisotropy in 
the multifractal) a greater difference in $p_{c_{ave}}$ will appear. 
Figure \ref{fig4} confirm this tendency for diverse values of $\rho$. 

Figure \ref{fig4} shows  $\bar p_c$ versus $\rho$. To obtain  $\bar p_c$ we  
make an average of $p_{c_{ave}}$ after the saturation process, it means, 
for $n \geq 8$. The  parameter  
 $\rho$ is indicated in the figure. 
These values are also shown in Table $I$.  The straight line 
in the figure is the linear fitting of the data.
 We focus our attention on two 
features of the figure: the general tendency of decreasing 
$p_c$ with $\rho$, 
and the anomalous case $(3,1)-Q_{mf}$. The main tendency of 
decreasing $p_{c}$ with $\rho$ we have discussed in connection 
with anisotropy.  We comment the anomalous situation  
of $(3,1)-Q_{mf}$ in relation to 
 topologic properties of $Q_{mf}$ which  are analyzed in 
what follows.

The object $Q_{mf}$ is build of a set of blocks $i$ with different areas,
 $A_i$, and number of neighbors, $\zeta_i$. The number  $\zeta_i$ is a central 
quantity in the study of topology.  
A topologic quantity, that is important in the investigation of the percolation 
threshold, is $\zeta_{ave}$ which is an average of $\zeta_i$  over 
 $Q_{mf}$. It means $\zeta_{ave} = \frac{\sum_{i} \zeta_i }{N} $ 
where $N=2^{2\>n}$ is the total 
number of blocks and the sum in the numerator is performed over the full 
multifractal.  Typically $\zeta_{ave}$ is a number that does not 
increase with  $n$, but attains a constant value after  
$n \approx 6$. We plot in Table $I$ this topologic quantity for $n=8$ 
and several values of $\rho$. 

\centerline{\it Table I}
\begin{tabbing}
  $(s,r)$      \hspace{0.6cm}        \= (1,1) \hspace{0.7cm} \= (4,3) \hspace{0.7cm}\=      (3,2) \hspace{0.7cm} \= (2,1) \hspace{0.7cm} \= (5,2)     \hspace{0.7cm} \=(3,1)      \hspace{0.7cm} \= (4,1)   \hspace{0.7cm} \= (5,1)\\
\hspace{0.2cm} $p_{c}$  \hspace{0.6cm} \= 0.5929 \hspace{0.4cm} \= 0.5262  \hspace{0.4cm} \= 0.5262  \hspace{0.4cm} \= 0.5256   \hspace{0.4cm} \=  0.5252   \hspace{0.4cm} \= 0.5253  \hspace{0.4cm} \= 0.5243   \hspace{0.5cm} \= 0.5241 \\
\hspace{0.2cm} $ \zeta_{ave} $   \hspace{0.6cm} \= 999 \hspace{0.6cm} \= 5.436 \hspace{0.7cm} \= 5.436  \hspace{0.6cm}  \= 5.434 \hspace{0.7cm} \= 5.436    \hspace{0.6cm} \= 5.426  \hspace{0.7cm} \= 5.436  \hspace{0.6cm} \= 5.436 \\

\end{tabbing}

We observe in the row of $ \zeta_{ave} $ that all these values are roughly the same, 
the exception corresponds to the case of the $(3,1)-Q_{mf}$. 
The conclusion we take is that the variation of the average 
number of neighbors causes the fluctuation in percolation threshold observed in 
figure \ref{fig4}. The $p_c$ 
of the $(3,1)-Q_{mf}$ is a little bit greater than the other cases 
because it has a smaller average number of neighbors. 
In other words, because each 
block of the $(3,1)-Q_{mf}$, in the average, has less neighbors, 
it percolates with more difficulty than 
the tendency among its group.

\section{4- Final Remarks}

To summarize we analyze in this work the role of the anisotropy and the 
average number of neighbors of $Q_{mf}$ on its 
percolation threshold, $p_c$. The multifractal object is 
composed by a set of blocks with different areas and number of neighbors. 
As  the parameter defining $Q_{mf}$,  $\rho$, goes to 
zero, the multifractal becomes more and more anisotropic. This 
anisotropy reflects in the percolation cluster creating an asymmetry 
between $p_c^e$ and $p_c^b$. 

The anisotropy of  $Q_{mf}$ is evident when we compare 
$p_c^e$ and $p_c^b$.  The observed 
curves of $p_c^e$ show a similar behavior, in contrast to 
the curves of  $p_c^b$. Actually, the measure of $p_c^b$ makes 
an average over both directions which erases any anisotropic 
effect of the percolation cluster. This erasing effect does not 
exist when we measure $p_c^e$. The anisotropy of  $Q_{mf}$ 
decreases with $\rho$ (with $\rho \rightarrow 0$ the blocks 
became more stretched).

A special case in our analysis is the  $(3,1)-Q_{mf}$. This 
case is singular compared to others analyzed cases. 
For a same number of blocks, 
the $(3,1)-Q_{mf}$ case  has less neighbors than the multifractal blocks  
corresponding to other values of $\rho$. 
This phenomenon is intrinsic to the topology 
of  $Q_{mf}$.  It implies that the $(3,1)-Q_{mf}$  
is less connected, and, as a consequence, it shows more difficulty to  
percolate. Therefore the case of the $(3,1)-Q_{mf}$ 
has a percolation threshold 
slighter greater then its neighbors in the $\rho$ sequence. 
In a future work we intend to 
study in more detail the effect of other topologic properties on 
the percolation properties.

\vspace{2cm}

The authors gratefully acknowledge the financial support of Conselho Nacional
de Desenvolvimento Cient{\'\i}fico e Tecnol{\'o}gico (CNPq)-Brazil, 
FINEP and CTPETRO.

\vspace{1cm}

\centerline{FIGURE LEGENDS}

\begin{figure}[ht]
\begin{center}
\caption{ The initial step, $n=1$, in the formation of $Q_{mf}$. 
In (a) a vertical line cut the
square in two pieces according to  $\rho$. 
Two horizontal lines sectioning the rectangles by
 the same ratio are depicted in (b). The underlying square lattice 
is depicted  with thick lines. }
\label{fig1}
\end{center}
\end{figure}
\begin{figure}[ht]
\begin{center}
\caption{The object $Q_{mf}$ for $\rho = \frac{1}{3}$ and $n=4$, we 
introduce a square inside the figure to help the visualization. }
\label{fig2}
\end{center}
\end{figure}
\begin{figure}[ht]
\begin{center}
\caption{In (a) we show $p_c$ versus $\frac{1}{L}$ for  
$\rho = \frac{2}{3}$ (solid line) and 
 $\rho = \frac{1}{4}$ (dashed line). The upper values correspond to $p_c^b$ 
 and the lower values to $p_c^e$; it is used $4 \leq n \leq 10$. 
In (b) we plot $p_{ave}$ 
versus $\frac{1}{L}$ for the same data. }
\label{fig3}
\end{center}
\end{figure}
\begin{figure}[ht]
\begin{center}
\caption{The  $\bar p_c$ versus $\rho$. 
The chosen values 
of $\rho$ are indicated in the figure. The straight line 
corresponds to the linear fitting.}
\label{fig4}
\end{center}
\end{figure}

\end{document}